\documentclass[reprint,superscriptaddress,amsmath,amssymb,aps,pra]{revtex4-1}

\usepackage{graphicx,epstopdf}
\usepackage{dcolumn}
\usepackage{bm}
\usepackage{hyperref}

\usepackage{multirow}
\usepackage{threeparttable}
\usepackage{color}
\usepackage{subfigure}
\usepackage{setspace}

\begin{document}
\title{Linear Optical Approach to Supersymmetric Dynamics}
\author{Yong-Tao Zhan}
\affiliation{CAS Key Laboratory of Quantum Information, University of Science and Technology of China, Hefei 230026, People's Republic of China}
\affiliation{Synergetic Innovation Center of Quantum Information and Quantum Physics, University of Science and Technology of China, Hefei 230026, People's Republic of China}
\author{Xiao-Ye Xu}
\email{xuxiaoye@ustc.edu.cn}
\affiliation{CAS Key Laboratory of Quantum Information, University of Science and Technology of China, Hefei 230026, People's Republic of China}
\affiliation{Synergetic Innovation Center of Quantum Information and Quantum Physics, University of Science and Technology of China, Hefei 230026, People's Republic of China}
\author{Qin-Qin Wang}
\affiliation{CAS Key Laboratory of Quantum Information, University of Science and Technology of China, Hefei 230026, People's Republic of China}
\affiliation{Synergetic Innovation Center of Quantum Information and Quantum Physics, University of Science and Technology of China, Hefei 230026, People's Republic of China}
\author{Wei-Wei Pan}
\affiliation{CAS Key Laboratory of Quantum Information, University of Science and Technology of China, Hefei 230026, People's Republic of China}
\affiliation{Synergetic Innovation Center of Quantum Information and Quantum Physics, University of Science and Technology of China, Hefei 230026, People's Republic of China}
\author{Munsif Jan}
\affiliation{CAS Key Laboratory of Quantum Information, University of Science and Technology of China, Hefei 230026, People's Republic of China}
\affiliation{Synergetic Innovation Center of Quantum Information and Quantum Physics, University of Science and Technology of China, Hefei 230026, People's Republic of China}
\author{Fu-Ming Chang}
\affiliation{CAS Key Laboratory of Quantum Information, University of Science and Technology of China, Hefei 230026, People's Republic of China}
\affiliation{Synergetic Innovation Center of Quantum Information and Quantum Physics, University of Science and Technology of China, Hefei 230026, People's Republic of China}
\author{Kai Sun}
\affiliation{CAS Key Laboratory of Quantum Information, University of Science and Technology of China, Hefei 230026, People's Republic of China}
\affiliation{Synergetic Innovation Center of Quantum Information and Quantum Physics, University of Science and Technology of China, Hefei 230026, People's Republic of China}
\author{Jin-Shi Xu}
\affiliation{CAS Key Laboratory of Quantum Information, University of Science and Technology of China, Hefei 230026, People's Republic of China}
\affiliation{Synergetic Innovation Center of Quantum Information and Quantum Physics, University of Science and Technology of China, Hefei 230026, People's Republic of China}
\author{Yong-Jian Han}
\email{smhan@ustc.edu.cn}
\affiliation{CAS Key Laboratory of Quantum Information, University of Science and Technology of China, Hefei 230026, People's Republic of China}
\affiliation{Synergetic Innovation Center of Quantum Information and Quantum Physics, University of Science and Technology of China, Hefei 230026, People's Republic of China}
\author{Chuan-Feng Li}
\email{cfli@ustc.edu.cn}
\affiliation{CAS Key Laboratory of Quantum Information, University of Science and Technology of China, Hefei 230026, People's Republic of China}
\affiliation{Synergetic Innovation Center of Quantum Information and Quantum Physics, University of Science and Technology of China, Hefei 230026, People's Republic of China}
\author{Guang-Can Guo}
\affiliation{CAS Key Laboratory of Quantum Information, University of Science and Technology of China, Hefei 230026, People's Republic of China}
\affiliation{Synergetic Innovation Center of Quantum Information and Quantum Physics, University of Science and Technology of China, Hefei 230026, People's Republic of China}

\date{\today}

\begin{abstract}
The concept of supersymmetry developed in particle physics has been applied to various fields of modern physics. In quantum mechanics, the supersymmetric systems refer to the systems involving two supersymmetric partner Hamiltonians, whose energy levels are degeneracy except one of the system has an extra ground state possibly, and the eigenstates of the partner systems can be mapped onto each other. Recently, an interferometric scheme has been proposed to show this relationship in ultracold atoms [Phys.\,Rev.\,A \textbf{96},\,043624\,(2017)]. Here this approach is generalized to linear optics for observing the supersymmetric dynamics with photons. The time evolution operator is simulated approximately via Suzuki-Trotter expansion with considering the realization of the kinetic and potential terms separately. The former is realized through the diffraction nature of light and the later is implemented using phase plate. Additionally, we propose an interferometric approach which can be implemented perfectly using amplitude alternator to realize the non-unitary operator. The numerical results show that our scheme is universal and can be realized with current technologies.
\end{abstract}

\maketitle

\section{Introduction}

In particle physics, to overcome the deficiencies in explaining the dark matter (energy) and the hierarchy problem via the standard model, theoretical physicists developed a concept called supersymmetry (SUSY)\,\cite{Dine2016}. Although SUSY can unify the weak, strong and electromagnetic couplings at high energy and give birth to the dark matter (energy), the existence of SUSY is still controversial due to lack of experimental proof\,\cite{Kane2001,Weinberg2005,Grodon2010,Nath2016}.  However, with the theoretical developments in nearly half a century, its conceptual structure has provided powerful mathematical tools (SUSY algebras), which have been applied to various fields apart from the high energy physics\,\cite{Binetruy2012}. One of such fields is the supersymmetric quantum mechanics\,\cite{Witten1981,Sukumar1985,Cooper1995,Gangopadhyaya2011}, which involves two supersymmetric partner Hamiltonians. This is a critical concept in SUSY, which originally describes the symmetry between the two basic particles fermions and bosons\,\cite{Haber2017}. Using this concept, a wide range of physical problems can be tackled out, e.g., the physics in stochastic dynamics\,\cite{Kinematic2016} and in the condensed matter\,\cite{Efetov1999}, new technologies can develop in other subjects, for example, the supersymmetric transformation optics\,\cite{Miri2013,Miri2014,Heinrich2014}.

Supersymmetric quantum mechanics establishes a relation between two partner Hamiltonians. Suppose one of the Hamiltonians $H^{(1)}$ can be factorized as $H^{(1)} = \hat{B}\hat{B}^\dag$, where $\hat{B}^\dag$ denotes the Hermitian conjugate of $\hat{B}$, its supersymmetric partner can then be defined as $H^{(2)} = \hat{B}^\dag\hat{B}$. We can easily find that the eigenenergies of the two systems with Hamiltonians given by $H^{(1)}$ and $H^{(2)}$ are the same (except  for  the  ground  state  of  one Hamiltonian  possibly), and their eigenstates can be mapped onto each other through operator $\hat{B}^\dag$ and $\hat{B}$. This relationship is a direct consequence of supersymmetric algebra, which is independent of the concerned subjects and can be directly applied to other fields.

\begin{figure}
  \centering
  \includegraphics[width=0.45\textwidth]{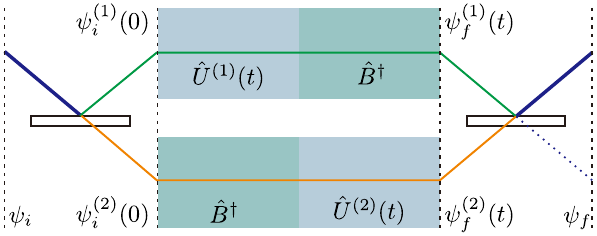}
  \caption{Interferometric approach to implement the supersysmmetric dynamics\,\cite{Lahrz2017}. The system, which is initialized in $\psi_i$, is split into two parts: one evolves the time evolution $\hat{U}^{(1)}(t)$ firstly and thereafter an operator $\hat{B}^\dag$ (green line); the other is diametrically opposite (the orange line). 
  The experimental setup of $\hat{U}(t)$ is plotted in FIG.2(an alignment of phase plates) and $\hat{B}^\dag$ in FIG.3(a Mach-Zehnder inteferometer).
  It is supposed that, if the two Hamiltonians governing the evolution of the two parts are superpartner, the final states $\psi_f^{(1)}(t)=\hat{B}^\dag \hat{U}^{(1)}(t) \psi_i$ and $\psi_f^{(2)}(t)=\hat{U}^{(2)}(t) \hat{B}^\dag \psi_i$ 
  are equal and a constructive interference should be observed after the second beam splitter.}\label{fig.MZ}
\end{figure}

Recently, Lahrz \emph{et al.} proposed a scheme for implementing the supersymmetric dynamics in ultracold atom systems\,\cite{Lahrz2017}. In their scheme, they considered verifying the translated version of the supersymmetric relationship between two partner Hamiltonians 
\begin{equation}\label{Eq.relation}
  \hat{B}^\dag \hat{U}^{(1)}(t) = \hat{U}^{(2)}(t)\hat{B}^\dag,
\end{equation}
where $\hat{U}^{(i)}(t)~(i\in\{1,2\})$ is the time evolution operator of the Hamiltonian $H^{(i)}$. The statement Eq.\ref{Eq.relation} holds for any time $t$, independently of the system's initial state. It is suggested to detect the SUSY via a Mach-Zehnder interference experiment. As shown in Fig.\ref{fig.MZ}, the system is initialized in state $\psi_i$ and then split to evolve along two paths corresponding to the supersymmetric partner Hamiltonians.
If Eq.\ref{Eq.relation} holds, a constructive interference is supposed to be observed for arbitrary evolution time $t$.

Although they have proposed to implement the supersymmetric dynamics in ultracold atom systems with current technology, their scheme has several deficiencies. First, the potential and the initial state are restricted to a specific form in their scheme, which makes the experiment results not convincing enough to verify the supersymmetric relation. Second, their approach in realizing the operator $\hat{B}^\dag$ (non-unitary) is simulating $(\hat{B}+\hat{B}^\dag)$ with a unitary operator. Although the first $\hat{B}^\dag$ can be implemented perfectly via setting the initial state to be a ground state, the fidelity of the second $\hat{B}^\dag$ operation can only reach $1/\sqrt{2}$.

In this article, we propose to extend their scheme in implementing and detecting the supersymmetric relation (Eq.\ref{Eq.relation}) into linear optics. We considered implementing the unitary time evolution operator $\hat{U}^{(i)}(t)$ and the non-unitary operator $B^\dag$ with currently available optical elements. Our scheme overcame the above difficulties. First, the system can be initialized in arbitrary state via the spatial light modulator (SLM) and the potential can be set to arbitrary form  by modifying phase plates. Second, we present a direct approach to implement $\hat{B}^\dag$ with a Mach-Zehnder interferometer, which can simulate the $\hat{B}^\dag$ operation with high fidelity.

The article is organized as follows. In Sec.II, we show how to implement operator $\hat{U}^{(i)}(t)$ and $\hat{B}^\dag$ in linear optics, which are two main problems in experimentally implementing supersymmetric dynamics. In Sec.III we set the initial state and potential to a specific form and numerically simulate the process of experiment. In addition, we analyze the precision of our realization. Finally, we present our conclusion in Sec.IV.

\begin{figure}
  \centering
  \includegraphics[width=0.48\textwidth]{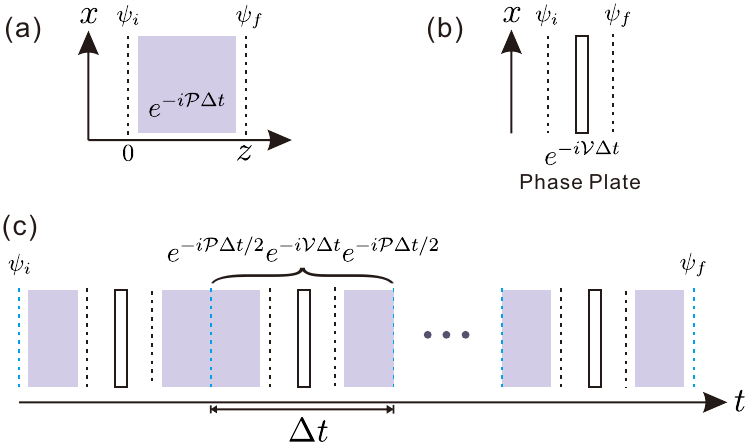}
  \caption{Implementation of time evolution operator. (a) depicts the diffraction of light in free space (shadow) as the simulation of the propagator of free particle. (b) shows a phase plate (rectangle) acting as a time evolution operator of a delta-kick potential. In (c), combination of (a) and (b) can realize a general time evolution operator for any time scale via the approximation of Suzuki-Trotter expansion (second order).}\label{fig.expsetup}
\end{figure}

\section{Implementing the time evolution operator in linear optics}
\subsection{Suzuki-Trotter expansion of the Time Evolution Operator}

Generally, the Hamiltonian of an isolated system contains two terms, i.e., kinetic energy $\mathcal{P}$ and potential energy $\mathcal{V}$ and the time evolution are described by the Schr\"odinger equation. However, solving such a differential equation is usually not an easy task even for a simple form of potential. For a time-independent scenario, the time evolution operator is in the form $\hat{U}(t) = \exp{[-i(\mathcal{P}+\mathcal{V})t]}$, which is also hard to solve for the noncommutative feature of the two terms. Analog quantum simulation\,\cite{Alan2012,Georgescu2014} of this time evolution needs to construct a controllable system with the same form of the Hamiltonian, which also faces challenges in the experiment.

To date, a variety of approximate formulas have been developed for calculating the time evolution operator\cite{Makri1989,Zagury2010,Argeri2014}. The leading one is the Suzuki-Trotter expansion, also called the exponential product formula\,\cite{Suzuki1976}. To the first and second order, it reads
\begin{equation}\label{Eq.first}
e^{t(\hat{A}+\hat{B})} = \lim_{n\rightarrow\infty}(e^{\hat{A} \Delta t}e^{\hat{B} \Delta t})^n
\end{equation}
and
\begin{equation}\label{Eq.second}
e^{t(\hat{A}+\hat{B})} = \lim_{n\rightarrow\infty}(e^{\hat{A}\Delta t/2}e^{\hat{B} \Delta t}e^{\hat{A}\Delta t/2})^n
\end{equation}
separately where $\hat{A}$ and $\hat{B}$ are arbitrary operators with $t=n \Delta t$. 
By applying the formula for the second order expansion (Eq.\,\ref{Eq.second}) to our scenario, 
we can decompose the time evolution operator $\hat{U}(t)$ into two terms, one only contains the kinetic term with the other only containing the potential term, which reads
\begin{equation}\label{Eq.teo}
\hat{U}(t) = \lim_{n\rightarrow\infty}(e^{-i\mathcal{P}\Delta t/2}e^{-i\mathcal{V}\Delta t}e^{-i\mathcal{P}\Delta t/2})^n.
\end{equation}
Instead of building a quantum simulation apparatus whose Hamiltonian is in the same form of our concerns, we reach the time evolution operator $\hat{U}(t)$ approximately following Eq.\,\ref{Eq.teo}. The kinetic and potential terms in a single step of the expansion can then be implemented separately and repeated for $n$ times, as shown in Fig.\,\ref{fig.expsetup}, which will be discussed in detail in the following subsection.

In the experiment, the number of expansion steps $n$ is finite, which cause a systematic error in this approach. Taking into account $n$ steps, the error after the time $t$ is $\epsilon = t^3/n^2$ for the second order expansion\cite{Dhand2014}. For a given time $t$, this systematic error can be reduced through improving the number of expansion steps. The precision of this realization is discussed in Sec.IV. 

\subsection{Diffraction as a free propagator}
In non-relativistic quantum mechanics, the time evolution of a system is depicted by the Schr\"odinger equation, whose fundamental solution can be given through Green's function, also known as the propagator\,\cite{Sakurai2011}. For a system with a Hamiltonian $H$, the propagator can be written as

\begin{equation}\label{Eq.propagator1}
    K(x,t;x',t') = \langle x|\hat{U}(t,t')|x'\rangle
\end{equation}
where $\hat{U}(t,t')$ represents the time evolution operator from time $t'$ to $t$ and $x(x')$ denotes the general coordinates. The term of kinetic energy in time evolution operator is known as the propagator of free particle, in 1D which is 
\begin{equation}\label{Eq.propagator2}
    K(x,t;x',0) = \sqrt{\frac{1}{2\pi i t}}\exp[i\frac{  (x-x')^2}{2 t}].
\end{equation}
 
For simplicity, we choose natural units $\hbar=m=1$. Consequently, the time evolution operator of a free particle ($\mathcal{V}=0$) can be implemented via the propagator in Eq.\ref{Eq.propagator2}. Here, we show that the small angle diffraction (propagation) of light in free space actually has the same form as the evolution of a free particle. 

In optics, the propagation of light follows the Huygens-Fresnel principle, which is stated as
\begin{equation}\label{Eq.propagation}
    E(x,y,z) = \frac{z}{i\lambda}\iint E(x',y',0)\frac{e^{ikr}}{r^2}\mathrm{d}x'\mathrm{d}y'.
\end{equation}
In a more simple and usable expression, Eq.\,\ref{Eq.propagation} can be approximated by the Fresnel diffraction integral, in 1D scenario which reads \begin{equation}\label{Eq.Fresnel}
    E(x,z) =\int F(x,z;x',0)E(x',0)\mathrm{d}x',
\end{equation}
where 
\begin{equation}\label{Eq.propagatorLight}
    F(x,z;x',0) = \exp(ikz) \sqrt{\frac{k}{2\pi i z}}\exp[i\frac{k (x-x')^2}{2z}].
\end{equation}
If we let $z/k=t$ according to the de Broglie's relation $p=k$ in Eq.\,\ref{Eq.propagatorLight}, the propagator $F(x,z;x',0)$ in Fresnel diffraction takes exactly the same form as the free propagator in quantum mechanics except for a global phase term. 

To derive the Fresnel diffraction formula Eq.\,\ref{Eq.Fresnel}, we need to apply the paraxial approximation ($r\approx z$) to Eq.\,\ref{Eq.propagation} and then replace the spherical secondary wavelets with parabolic ones, which is known as the Fresnel approximation ($r\approx z+(x-x')^2/2z$). In the range of a small angle diffraction, i.e., $\rho^2/z^2\ll1$ with $\rho$ denoting the characteristic size of the light spot, the Fresnel approximation is equivalent to the paraxial approximation\,\cite{Goodman2017}. Therefore, the kinetic term in the time evolution operator $\hat{U}^{(i)}(t)$ can be simulated through the small angles diffraction of light in free space.  

\begin{figure}
  \centering
  \includegraphics[width=0.45\textwidth]{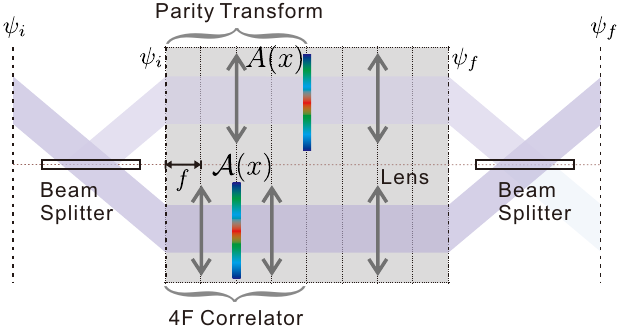}
  \caption{Mach-Zehnder interferometer with intensity modulator for implementing the non-Hermitian operator $\hat{B}^\dag$. In the upper arm, two lens acting as parity transforms are employed for light propagation without diffraction. An position dependent intensity modulator is inserted for implementing the term $\alpha W(x)$ in the operator. In the lower arm, we firstly adopt an 4f system with an intensity modulator inserted in the second focal point for simulating the operation in momentum space. Then we perform a parity transform with another lens. Combination of the two arms can give the target wave function.}\label{fig.Bdag}
\end{figure}

\subsection{Phase plate as a delta-kick potential}
Another term in the Suzuki-Trotter expansion of the time evolution operator is the potential, which will result in a pure phase shift $\exp{[-i V(x)\Delta t]}$ to the wave function after an evolution time $\Delta t$. Current technologies in adaptive optics can control the wave front arbitrarily with the phase plate or the spatial light modulator (a programmable version of the former)\,\cite{Tyson2015}. The phase plate can adjust the phase of the light wave front depending on the spatial position without changing its intensity distribution, which is equivalent to multiplying the wave function $\psi(x)$ with a position dependent phase term, i.e., $\exp{[-i\phi(x)]\psi(x)}$. 

In fact, the phase plate provides an implementation for the time evolution with a delta-kick potential, where a time independent potential $\phi(x)$ appears suddenly in one moment and then immediately disappear. The Hamiltonian for a system with the delta-kick potential takes the form
\begin{equation}
    -\frac{1}{2}\frac{\partial^2}{\partial x^2} +\phi(x)\delta(t),
\end{equation}
where $\delta(t)$ is the Dirac delta function. The corresponding time evolution operator takes the form of a pure phase shift $\hat{U} = \exp[-i\phi(x)]$. If we set $\phi(x) = V(x)\Delta t$, a phase plate can realize the potential term in the Suzuki-Trotter expansion (Eq.\ref{Eq.teo}), create a phase shift to the wave function. Here, the ``delta-kick" means the interaction time for the phase plate should be very short, so that the diffraction inside the phase plate can be ignored (the phase plate should be as thin as possible). In a simple word, the whole time evolution operator $\hat{U}^{(i)}(t)$ can be simulated approximately with an array composed of a free space diffraction (zero potential), a phase plate (delta-kick potential) and another free space diffraction in order, as shown in Fig.\,\ref{fig.expsetup}.

In the scheme proposed by Lahrz \emph{et al.}, the time evolution operator in atom systems is realized by creating the potential with detuned Gaussian beam which are rich in creating both p- and s-barrier. In this way, the form of the potential is restricted to be Gaussian-like. Our scheme, in contrast, can realize time evolution operators in any type of potentials. If we want to use another form of potential, we just need to change the phase modulation in the phase plate.

\subsection{Interferometer approach for non-unitary operator}
In this section, we discuss the realization of another important operator in the supersymmetric relationship, i.e., $\hat{B}^\dag = [-\partial_x +W(x)]/\sqrt{2}$. In the scheme of Lahrz \emph{et al.}, they proposed to realize $\hat{B} + \hat{B}^\dag$ instead of $\hat{B}^\dag$. The former can be approximated via the time evolution operator of a shaking process with harmonic potential. In their first implementation of $\hat{B}^\dag$, the initial state is set to be a ground state, as a result, $\hat{B} + \hat{B}^\dag$ degenerates to $\hat{B}^\dag$. However, the second $\hat{B}^\dag$ should be applied to a general state, $\hat{U}^{(1)}(t)\psi_i$, where the term of $\hat{B}$ will not degenerate. The simulation procedure can only reach a fidelity of $1/\sqrt{2}$. Here we show that $\hat{B}^\dag$ can be implemented with high fidelity in our setup.

As shown in Fig.\,\ref{fig.Bdag}, we propose to implement $\hat{B}^\dag$ via a Mach-Zehnder interferometer, one arm for the term $\partial_x$ and the other for $W(x)$. As $\hat{B}^\dag$ is non-unitary, non-unitary elements should be included, e.g., the position dependent intensity modulators. In optics, the lens can perform a Fourier transform to the wave front, that is to say, the wave function in momentum space can be modulated by lens and amplitude modulators. In the lower arm of the interferometer, we adopt a 4f system to realize the term in $\hat{B}^\dag$ involving the momentum.  We insert an intensity modulator $\mathcal{A}(x) = \alpha' kx/f$ at the focal point of the 4f system, where $\alpha'$ is a parameter to ensure $|\alpha' kx/f|<1$ and $f$ is the focal length of the lens. Minus $\mathcal{A}(x)$ means an additional $\pi$ phase should be added. After performing a parity inversion by using another lens with the same focal length, the wave function in this arm is changed to $-\partial_x \psi_i$ with multiplying a constant factor $\alpha'$. In the upper arm, for compensating the total length and avoiding the diffraction at this stage, we adopt two lenses to perform parity inversion and insert an intensity modulator $A(x) = \alpha W(-x)$ in the middle of the two lens (as shown in Fig.\,\ref{fig.Bdag}). $\alpha$ is a constant for renormalization and minus $A(x)$ means an additional $\pi$ phase should be added. The final wave function in this arm reads $\alpha W(x)\psi_i$.  
Finally, the combination of the two arms can give the complete realization of $\hat{B}^\dag$ with setting $\alpha = \alpha'$. Our approach to the non-unitary operator $\hat{B}^\dag$ can reach high fidelity, which is shown in the following numerical simulations. 

\section{Numerically Simulated Results}

\begin{figure}
  \centering
  \includegraphics[width=0.4\textwidth]{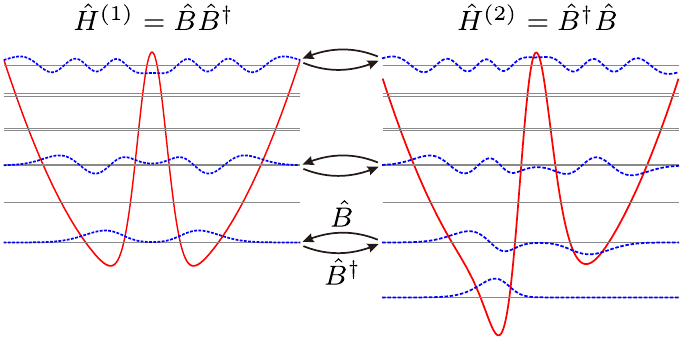}
  \caption{Supersymmetric potentials (red solid lines) and their energy levels (black solid lines). We can clearly see that the spectra of the two potentials coincide with only an exception of the ground state (unbroken symmetry). 
  The wave packet split into two packets during the evolution because of the existence of the barrier. }\label{fig.energylevel}
\end{figure}

As stated in the previous section, our scheme can be applied to any type of superpotential with adjusting the phase modulation of the phase plate. The initial state can also be prepared in any form with wave front modulators. Here, for the sake of better understanding, we consider the same scenario as reported in Ref.\,\cite{Lahrz2017},  where the superpotential reads 
\begin{equation}
    W(x) =\sqrt{\omega}\{x/x_0 + A\exp[-x^2/(4\sigma^2)]\}.
\end{equation}
Thus the potentials $V^{(1)}(x)$ and $V^{(2)}(x)$ in the  supersymmetric partner Hamiltonians take the form
\begin{eqnarray}
  V^{(i)}(x)=&&\frac{\omega^2x^2}{2}\mp\frac{\omega}{2}+\frac{\omega A^2}{2}\exp{(-\frac{x^2}{2\sigma^2})}\nonumber\\
  &&\frac{2\omega Ax}{x_0}(1\mp\frac{x_0^2}{4\sigma^2})\exp{(-\frac{x^2}{4\sigma^2})},
\end{eqnarray}
which contains two terms, the harmonic potential $V_{\text{osc}}=\omega^2 x^2/2$ and the additional potential $V_{\text{loc}}^{(i)}$. The parameter $A$ was set to $\sqrt{26}$. In the potential $V_{\text{osc}}$, $\omega$ is the trap frequency and $x_0 = \sqrt{1/\omega}$ stands for the harmonic oscillator length. The term of additional potential acts as the barrier, which falls off on a length scale $\sigma=x_0/2$. The structure of the energy level for the two potentials are plotted in Fig.\ref{fig.energylevel}, where the supersymmetric relationship is shown clearly.

\begin{figure}
  \centering
  \includegraphics[width=0.5\textwidth]{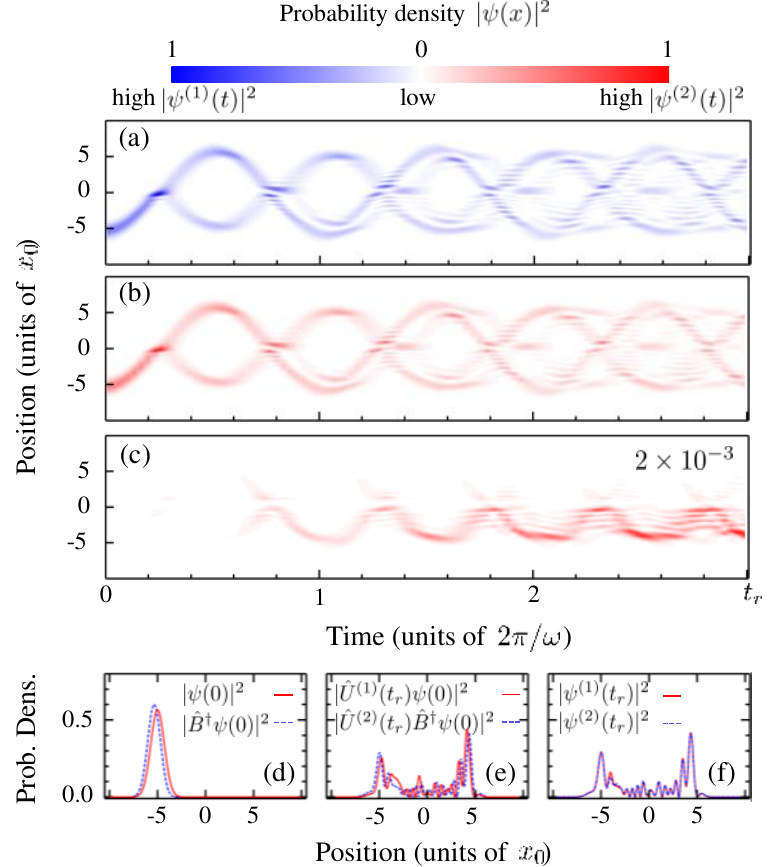}
  \caption{The normalized probability densities are plotted in (a) for $|\psi_f^{(1)}(t)|^2$ and in (b) for $|\psi_f^{(2)}(t)|^2$. In (c) we show the  interference pattern between the two paths $| \psi_f^{(1)}(t)- \psi_f^{(2)}(t)|^2$. In (d) we show the normalized probability densities of $|\psi(0)|^2$ and $| \hat{B}^\dag\psi(0)|^2$ before the time evolution. In (e) we show those at time $t_r=3 T$. 
  In (f), the probability density of the upper path after performing the second $\hat{B}^\dag$ operation coincides with the lower path very well, which indicates the existence of supersymmetry.}\label{fig.evolution}
\end{figure}

For optical simulation, it is more convenient to take the natural length and energy scales, where the energy is measured in units of $\omega$ and the distance is measured in units of its characteristic length $x_0$. The system is initialized to be a shifted Gaussian wave function $\psi(0) = (\pi x_0^2)^{-1/4}\exp{[-(x-5x_0)^2/(2x_0^2)]}$. For optical simulation, we choose the photon's wavelength $\lambda=532$\,nm and the characteristic length $x_0=1$\,mm. The evolution time of a single step in the Suzuki-Trotter expansion is set to $\Delta t = T/60$ with $T = 2\pi/\omega$ standing for the period, which results in a corresponding  propagation distance $z=1.237$\,m for diffraction between phase plates. Through simulating the experiment we plot the normalized probability densities of the final wave functions $\psi_f^{(1)}(t) = \hat{B}^\dag  \hat{U}^{(1)}(t)\psi(0)$ and $\psi_f^{(2)}(t) =  \hat{U}^{(2)}(t)\hat{B}^\dag\psi(0)$  with $t$ ranging from $0$ to $t_r=3T$ in Fig.\ref{fig.evolution}(a) and (b). In (c), we show the deviation between the two wave functions, i.e., the probability intensity of $|\psi_f^{(1)}(t) - \psi_f^{(2)}(t) |^2$. It is obvious that the deviation is relatively small ($~10^{-3}$), indicating the existence of supersymmetry.

For further clarification, we show the probability density distribution for different stages during the time evolution in Fig.\ref{fig.evolution}(d-f). The initial density distributions $|\psi(0)|^2$ and $|\hat{B}^\dag\psi(0)|^2$ are depicted in Fig.\ref{fig.evolution}(d). $\psi(0)$ is a shifted ground state with its centre at $-5x_0$. In Fig.\ref{fig.evolution}(e), we show the probability distributions of the two states inside the two arms before the second $\hat{B}^\dag$ is applied, which are divergent clearly. After performing the second $\hat{B}^\dag$ to the state $\hat{U}^{(1)}(t_r)\psi(0)$ in the upper arm, the two final states match each other very well, which can be seen clearly from their probability distributions as shown in Fig.\ref{fig.evolution}(f). We have also estimated the fidelity between these two states, which reads 0.9973. The perfect overlap of these two final states clearly shows the existence of supersymmetry again.

\begin{figure}
  \centering
  \includegraphics[width=0.42\textwidth]{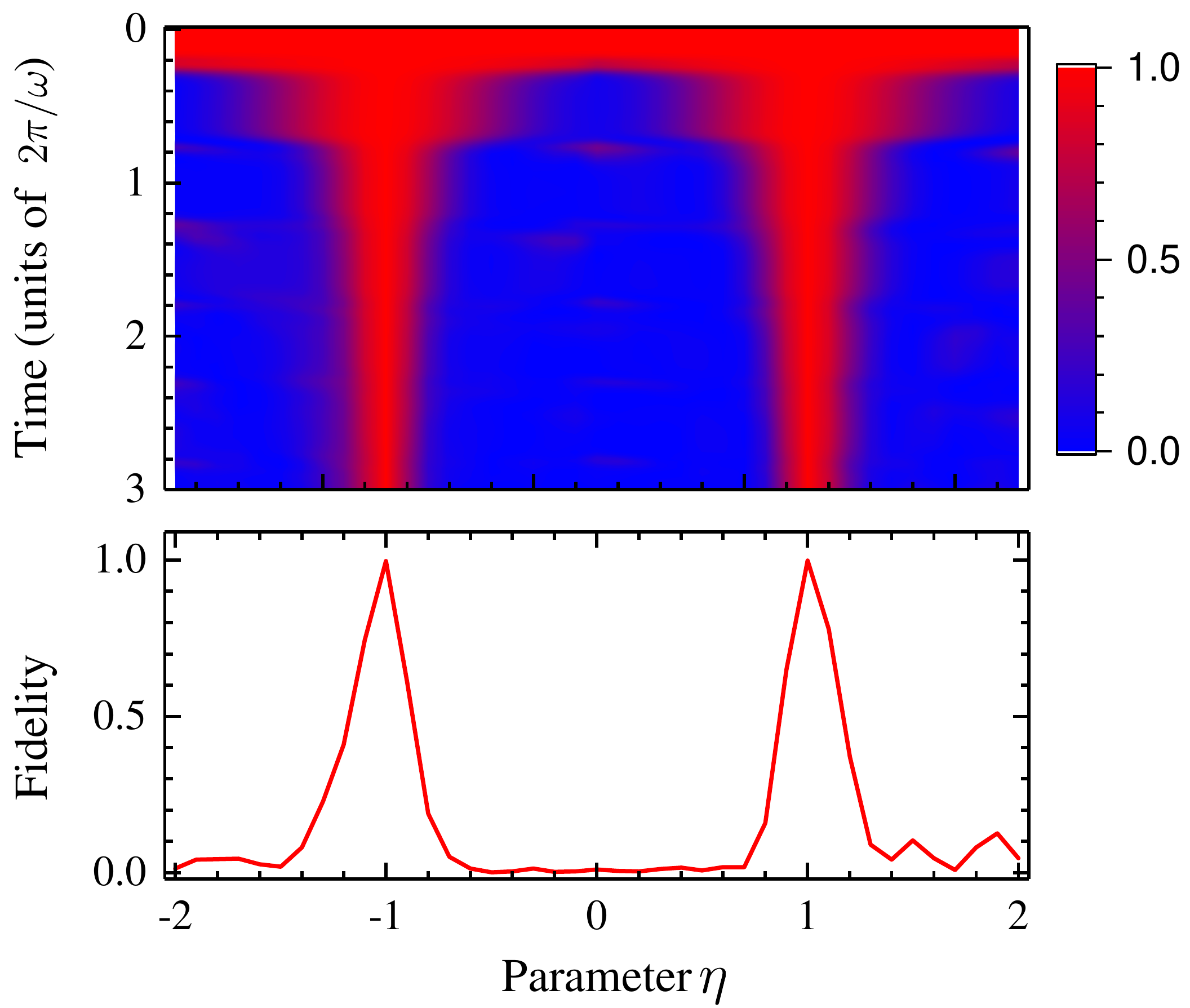}
  \caption{We show the fidelity between $\hat{B}^\dag \hat{U}^{(1)}(t)\psi(0)$ and $\hat{U}_\eta (t)\hat{B}^\dag\psi(0)$ in the upper panel with $t$ range from $0$ to $3 T$. The fidelity at $t_r = 3 T$ is shown in the bottom. The peak-points at $\eta=\pm 1$ demonstrate the existence of supersymmetry. }\label{fig.eta}
\end{figure}

The family of potentials characterized in Ref.\,\cite{Lahrz2017}  for further comparison is as follow
\begin{equation}
    V_\eta(x)= V_\text{osc}(x)+\frac{\omega A^2}{2}\exp{(-\frac{2x^2}{x_0^2})}+\frac{2\eta\omega Ax}{x_0}\exp{(-\frac{x^2}{x_0^2})}
\end{equation}
which were parameterized by $\eta$. The previous potentials then correspond to two special cases $\eta = 0$ for $V^{(1)}(x)$ and $\eta = 1$ for $V^{(2)}(x)$ with neglecting the constant term. As discussed in the previous section, a high fidelity (up to 1) between the final wave functions involving in the evolution governed by the potential with $\eta = 1$ and $\eta=0$, can identify the existence of supersymmetry. We depict the numerically simulated fidelity between $\hat{B}^\dag\hat{U}^{(1)}(t)\psi(0)$ and $\hat{U}_\eta(t)\hat{B}^\dag\psi(0)$ varying with time and $\eta$ in the upper panel of Fig.\ref{fig.eta}. In the lower figure, we show the fidelity at time $t_r$ varying with $\eta$. We can see that for any evolution time, the fidelity peaks to $1$ when $\eta = \pm 1$, which clearly shows the existence of supersymmetry ($V_{\eta = -1}(x) $ is also a superpartner potential of $V_{\eta = 0}(x) $).

\begin{figure}
  \centering
  \includegraphics[width=0.42\textwidth]{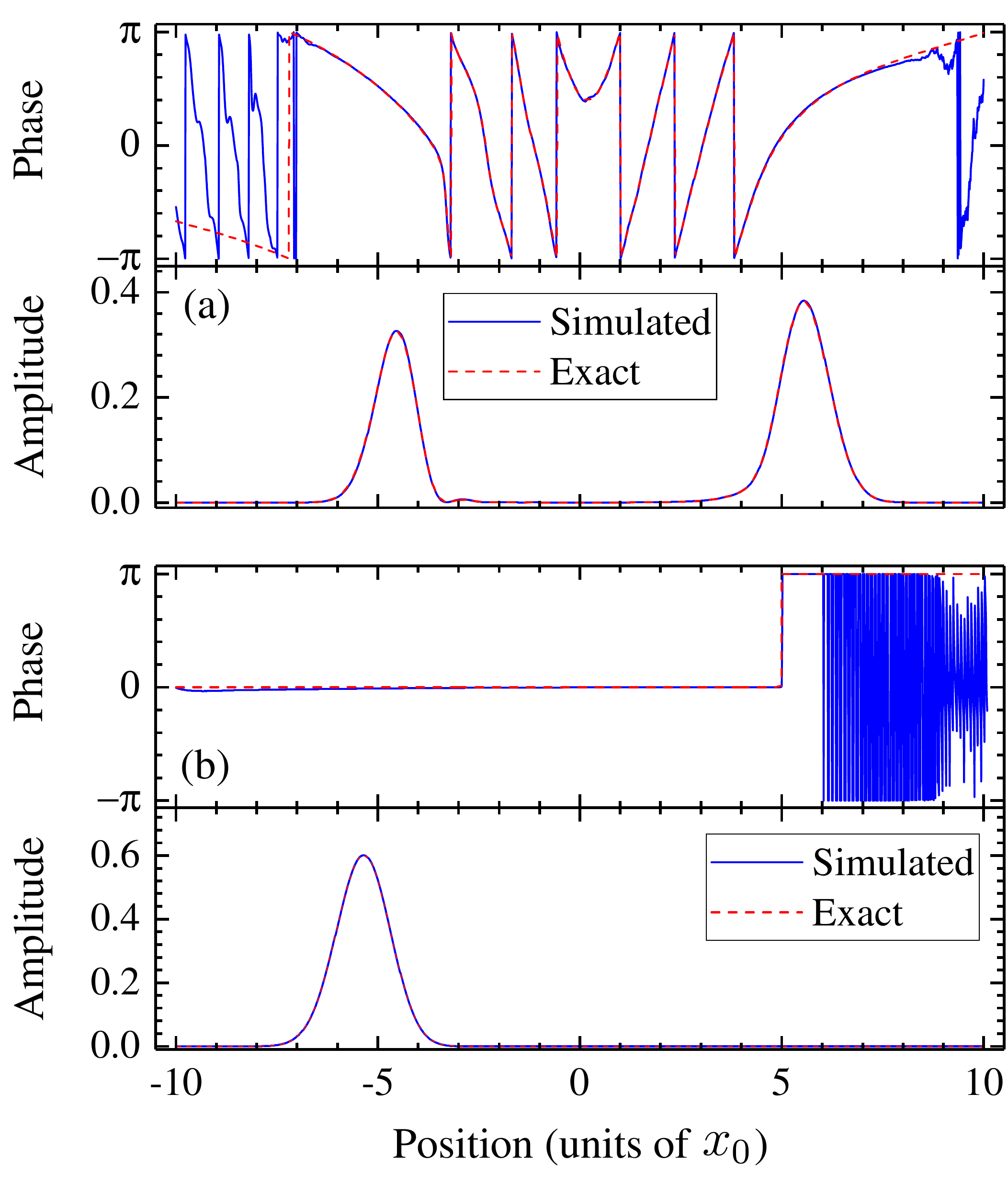}
  \caption{In (a) we set the initial state to $\hat{B}^\dag \psi(0)$ and compare the final states $\hat{U}^{(2)}(T/2)\hat{B}^\dag \psi(0)$ which are calculated through solving the Schr\"odinger equation (red dashed line) and simulating the experimental scheme (blue solid line). In (b) we show the final states $\hat{B}^\dag\psi(0)$ which are obtained through algebraic calculation and simulation of the experimental scheme. In both (a) and (b) the two wave functions overlap very well except some low-probability regions $<10^{-6}$ because of the numerical precision, where the phases diverse the exact values. In (a) the fidelity of the two states reaches 0.99976. In (b) the error rate is lower than $10^{-5}$.}\label{fig.estimationtwo}
\end{figure}

Our scheme is based on an approximate method, i.e., the Suzuki-Trotter expansion. The separated terms in the expansion can be implemented with high fidelity using current available optical elements. So it is necessary to estimate the fidelity of the time evolution operator.
Higher order approximation in Suzuki-Trotter expansion can provide higher operation fidelity, which is also more difficult to realize in the experiment. Here we adopt the second order approximation, which is easy to realize and can provide enough operation fidelity. Considering the time evolution of the second subsystem, i.e., the time evolution along the lower arm, the initial state is $\hat{B}^\dag\psi(0)$ and the final state after half a period of evolution, i.e. at time $T/2$ can be calculated by performing an operator $\hat{U}^{(2)}(T/2)$ on the initial state. We need to repeat the calculation 30 cycles for the single step time evolution$(\Delta t = T/60$) to get the final state at $T/2$. We depict both its amplitude and phase by blue solid lines in Fig.\ref{fig.estimationtwo}(a) with the exact results from the solution of the system's Schr\"odinger equation shown by red dashed lines. It is clearly shown that both simulated and exact results are in good agreement except for the region where the probability is lower than $10^{-6}$. We compare the simulated and exact final states quantitatively by the fidelity, which reads 0.9998 for this time scale. This fidelity can also be improved via increasing the number of steps, i.e., dividing the time into shorter time slices, which requires shortening the diffraction length, provided the condition for the Fresnel approximation is fulfilled. In our simulation, the characteristic size of the light spot is around $10x_0$, therefore the condition of small angles of diffraction is fulfilled in our scenario ($\sim10^{-4}$). In our consideration, the small angles approximation is completely fulfilled, as the total spot size is $\rho\sim10$\,mm and the diffraction length $z\sim10^3$\,mm, then $\rho^2/z^2\sim10^{-4}\ll 1$. Additionally, we can squarely reduce the system size $z$ via reducing the characteristic length $x_0$ according to the constraints $z = 2\pi N x_0^2/\lambda$ ($N$ stands for the time interval $\Delta t$ in the units of $1/\omega$). The paraxial condition should be fulfilled after reducing the system size.

Further, we estimate the fidelity in case of non-unitary $\hat{B}^\dag$ for our current setup. In the numerical simulation, we take the lens as ideal and its effect is like a position dependent phase gate on the wave front, that is multiplying $\exp[-ik x^2/(2f)]$ (with $f$ corresponding to the focus length of the lens) to the wave function in the clear aperture $x\in[-10x_0,10x_0]$. The focal length of the lens is set to $f=0.8$\,m. In Fig.\ref{fig.estimationtwo}(b), we show both the amplitude and phase of the final state after acting the operator $\hat{B}^\dag$ on the initial state $\psi(0)$. It is clearly shown that our scheme can provide high fidelity (error rate lower than $10^{-5}$) for implementing $\hat{B}^\dag$. Our method needs the paraxial condition $f^2/\rho^2\gg 1$ which equals $6.4\times10^3$ in our numerical simulation. The further simulation shows that the error rate can be lower than $10^{-3}$ if it is bigger than $2.5\times10^3$. Additionally, we need to emphasize that our approach to $\hat{B}^\dag$ is independent of the initial state. 

\section{Discussion and Conclusion}
In this paper, we propose to extend the interferometer scheme for detecting the supersymmetric dynamics in Ref.\cite{Lahrz2017} into linear optics. It is shown in numerically simulated results that our protocol provides good remarks of the supersymmetric dynamics of the superpartner via the destructive interference. The full wave function of a single photon can be directly measured through a technology recently developed from the concept of  weak measurement and weak value\,\cite{Lundeen2011}. It is important to note that the interferometer approach is not necessary for our scheme. Both the final wave functions along the up and bottom arms can be fully reconstructed and then the supersymmetric relationship can be verified directly via calculating the fidelity between these two final wave functions. Such an alternative method can avoid the challenge of stabilizing the interferometer and prove more feasible for experiment with current technologies.

\begin{acknowledgments}
This work was supported by National Key Research and Development Program of China (Nos.\,2017YFA0304100, 2016YFA0302700), the National Natural Science Foundation of China (Nos.\,11474267, 61327901, 11774335, 61322506), Key Research Program of Frontier Sciences, CAS (No.\,QYZDY-SSW-SLH003), the Fundamental Research Funds for the Central Universities (No.\,WK2470000026), the National Postdoctoral Program for Innovative Talents (No.\,BX201600146), China Postdoctoral Science Foundation (No.\,2017M612073), and Anhui Initiative in Quantum Information Technologies (Grant No. AHY060300).
\end{acknowledgments}

\bibliography{references}
\end{document}